# Growth and Optical Properties Investigation of Pure and Al-doped $SnO_2$ Nanostructures by Sol-Gel Method

*Ali Reza Razeghizadeh[1,*+], Lila Zalaghi[1], Iraj Kazeminezhad[2], Vahdat Rafee[1]*

[1,*]*Department of Physics, Faculty of Science, Payamenoor University, Tehran P.O.BOX19395-3697, Iran*
[2]*Department of Physics, Faculty of Science, Shahid Chamran University of Ahvaz, Iran*

ABSTRACT: *$SnO_2$ nanoparticles with different percentage of Al (5%, 15%, and25%) were synthesized by sol-gel method. The structure and nature of nanoparticles are determined by of X-ray diffraction analysis. Also morphology of the samples is evaluated by SEM. Moreover, the optical properties of the samples are investigated with UV-Visible and FT-IR. The XRD patterns are indicated that all samples and incorporation aluminum ions into the $SnO_2$ lattice have tetragonal rutile structure. The crystalline size of nanoparticles is decreased with increasing the Al percentage. The SEM results confirmed that the size of nanoparticles decreases with increasing the Al percentage. Also, FT-IR and UV-Visible results showed that the optical band gap of nanoparticles increases with the increasing the Al percentage. Finally, we have used the EDX analysis to study the chemical composition of the products. Pure tin and oxygen have been observed. The doped samples showed the existence of Al atoms in the samples of the crystal structure of $SnO_2$.*

## INTRODUCTION

Semiconductors are one of the most interesting and most useful solids. They have been investigated many times because of their flexibility, electricity and optical features. $SnO_2$ is one of these semiconductors. Stoichiometric $SnO_2$ is a good insulator because it has little carriers in this circumstance, but non-stoichiometric $SnO_2$ is transparent and conductive because it has many charge carriers in this circumstance. Also, it has a direct optical band gap of about $(3.8 - 4.3eV)$ and an indirect band gap of about $(2.7 - 3.1eV)$ with 80% transparency in the visible rang [1-3]. It is suitable for use in gas sensors [4-6], solar cell, and optoelectronic devices and photo catalysts in a wide range [7-9] because it has a wide band gap and unique electronic and optical properties.

* ***To whom correspondence should be addressed.***
**+ E-mail**: *razeghizadeh@yahoo.com*

One of the most important parameters of the semiconductors is their energy band gap. This parameter affects many of electrical and optical properties of semiconductors. The energy band gap will be changed according to changes in temperature, pressure and size of the particles. Therefor this parameter can show new properties of the semiconductor [10].

There are many different methods to synthesize nanoparticles of $SnO_2$ such as sedimentation, thermal solvent, micro emulsions, sol-gel, hydrothermal, spray-pyrolysis, and the non-aqueous methods [2, 3, 11].

We choose sol-gel method because this method is cheap; it does not need any complicated instruments and homogeneous particles can be produced in this method [12, 13].

The change in optical and structural properties of $SnO_2$ with changing temperature rate was investigated earlier [14]. In other related works, the optical and structural properties of $SnO_2$ nano films were studied [15]. Also the structure of $SnO_2$ nanopowders was studied [16]. Synthesizing the nanoparticles of $SnO_2$ with ultrasonic was investigated [17]. The structural and optical properties of $SnO_2$ have been studied by other research groups [18].

Doping the $SnO_2$ nano films with antimony [19] and investigation of their photoluminescence [20] are other examples of past related researches.

For the first time we will report a sol-gel method with a different Al percentage to fabricate the $SnO_2$ nanoparticles. We present the preparation method of the $SnO_2$ nanoparticles. Moreover, we study the structural properties and surface morphology of the $SnO_2$ and Al doped nanoparticles by X-ray diffraction (XRD) and scanning electron microscopy (SEM). One of the most common tools in studying the optical properties of the substance and calculating their energy band gap is the transmission, absorption, and reflection spectrum. In this work we used the absorption spectrum. For measuring the optical properties we use the UV-vis spectrophotometer in the wavelength range of 200-1500nm.

## EXPERIMENTAL METHOD

*Synthesis $SnO_2$ nanostructures pure and doped with different amounts of Al:*

At first, we shed 2.25g $SnCl_2.2HO_2$ ($SnCl_2.2HO_2$ with molar mass 225.63g/mol and purity 98%and made by Merck Company) and 100 ml pure ethanol (Ethanol $C_2H_5OH$ with molar mass46.07g/mol purity 98% and made by Merck Company.) into the beaker and mix them using a magnetic mixer (Magnetic magnet stirs Genway 1000 model (Hot Plate).) to get 0.1 molar solution. Then, we add 0.0667, 0.2001 and 0.33335 $(g)$ of $AlCl_3$ ($AlCl_3$ with molar mass 133.34g/mol and purity 98% and made by Merck Company.) to 0.1 molar solutions little by little to get the Beaker to the setting refluxing and turn one heater to reach them $80°C$. In order to prevent evaporation, we used cold water flow to make the solution be refluxed for 1 hour.

At the end, we get clear Sol. We took the Sol away from humidity to complete the aging process for 24 hours.

After getting the Sol, we put it in the oven on $100°C$ (Electric oven with a maximum operating temperature $250°C$, Heraeus model.) to become dry and turned to a white Gel.

Then we get different solutions of Al percent, 5%, 15% and 25% respectively.

Finally dried Gel have been grained and heated to $500°C$ (Electric furnace with a maximum operating temperature $1100°C$ Lenton, thermal design model.) for 1 hour to get pure and different Al percent doped nanopowders of $SnO_2$.

The quantity of $AlCl_3$ to get desired uses of flowing formula:

$$\left(\frac{Al}{Sn}\right)\% = \frac{(AlCl_3)\times\frac{1}{133.34}(gr)}{(SnCl_2\cdot 2H_2O)\frac{1}{225.63}(gr)} \qquad (1)$$

RESULT AND DISCUSSION

*Structure properties*

*XRD Analysis*

The X-ray diffraction (X-pert model of Philips company, Seifert3003)) diffraction measurement was carried out using an X-Pert Philips, in 2$\theta$ range $20° - 80°$ using Cu K$\alpha$ radiation of wavelength $\lambda$ = 1.5406 Å operating. Fig.1 shows the peaks of pure and 5%, 15%, 25% Al doped $SnO_2$ powders. The peaks in the spectra are identified as originating from reflections from the (110), (101), (211), and (301). These show that the $SnO_2$ particles are in the tetragonal phase.

Also it was observed that increasing Al percentage in the powder will cause the intensity of XRD pattern to be decreased. Thus, increasing the Al percentage in the nanoparticles will cause them to go to an amorphous state.

The crystal size calculated from the Scherrer formula [21].

$$D = \frac{k\lambda}{\beta\cos\theta} \qquad (2)$$

Where $D$ is the average crystallite size, $\lambda$ is the applied X-ray wavelength, and $k = 0.98$ which is a constant, $\theta$ is the diffraction angle in degree and $\beta$ is the full width at half maximum (FWHM) of the diffraction peak observed in radians.

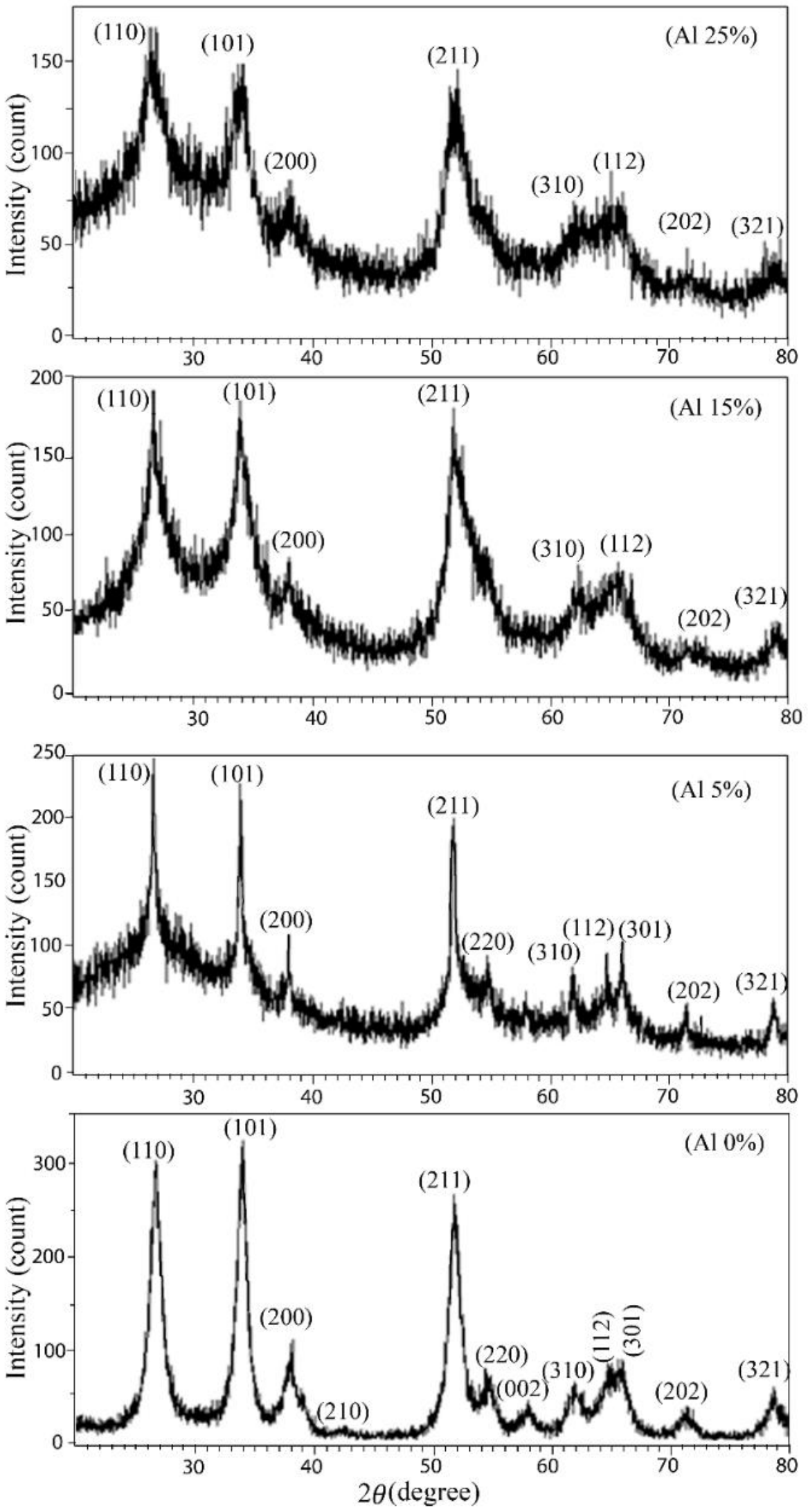

Fig. 1 XRD pattern of $SnO_2$ Nanopowders pure, and different Al-doped

Table 1 shows the crystallite size of pure and 5%, 15%, 25% Al doped of the powders are 11.7, 8.8, 7.7 and 7.2 nm, respectively.

The results show that the crystalline size of nanoparticles decreases with increasing Al percent of the sol. The position of the peaks in the pure and Al doped of the XRD pattern slightly shifted by changes in Al percent of the sol.

Table. 1 The crystallite size of samples with difference Al content

| Al –doped | Crystallite size (nm) |
|---|---|
| 0% | 11.7 |
| 5% | 8.8 |
| 15% | 7.7 |
| 25% | 7.2 |

*SEM Analysis*

Fig. 2 shows the SEM analysis (Scanning electron microscopy (SEM: S4160 model by the Hitachi company).) of surface morphology of nanoparticles, deposited on glass substrates for pure and 5%, 15%, 25% Al-dopes.

The average grain size is 18nm for pure and 16, 15 and 14nm for 5%, 15% and 25% Al doped nanoparticles, respectively. The images show that the grain size of the particles decreases when the percent of Al in the products increases. As it can be seen some of the grains agglomerate and make bigger clusters.

Fig. 3 and Table. 2 show the average nanoparticles size of SEM with difference Al percentage.

Table. 2 The average nanoparticles size of samples with difference Al content

| Al –doped | Average particle size (nm) |
|---|---|
| 0% | 18 |
| 5% | 16 |
| 15% | 15 |
| 25% | 14 |

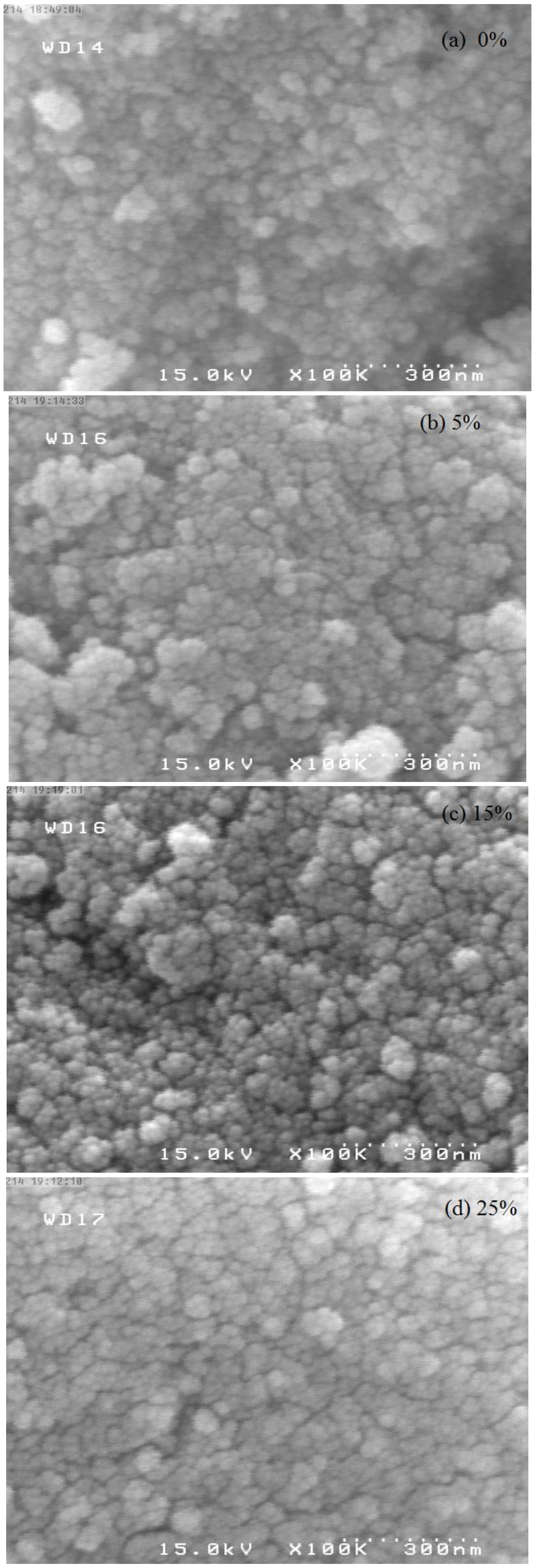

Fig. 2 SEM images of different Al-doped $SnO_2$ nanoparticles.

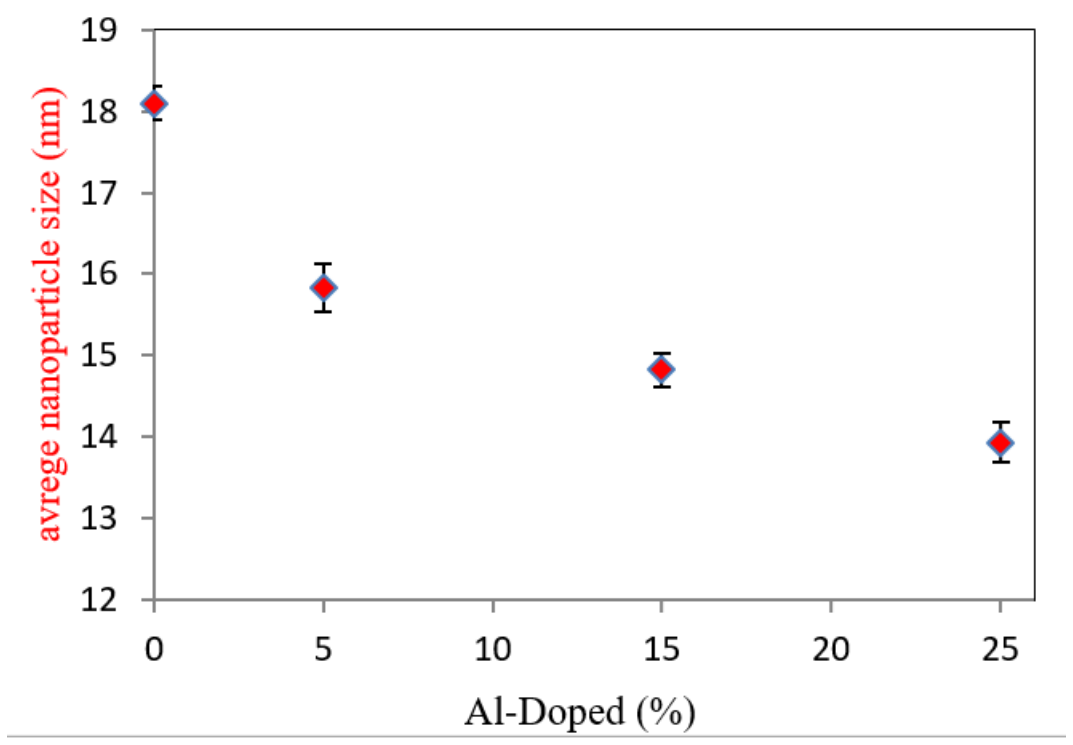

Fig.3Average size of nanoparticles with difference Al-doped

According to the SEM images, it is clear that increasing Al percentage will cause decreases in the size of nanoparticles, because tin ions have been replaced by Al ions on the structure. It is reasonable because the size of Al atom is smaller than the size of tin. These results are in complete agreement with XRD data.

*Optical Properties*

*UV-visible Analysis*

The optical properties were studied by UV-vis spectrophotometer (UV-Visible: (Varian-Cary5000 scan model spectrophotometer)). Fig. 4 shows the optical transmittance spectra of $SnO_2$ nanoparticles prepared at the different percentages of Al: 0%, 5%, 15% and 25%. We have considered photon wavelength range of 200--500*nm*. We find that nanoparticles have a high transmittance in the wider wavelengths.

There are three regions in the Fig.4 at transmission curve. In the first region (240 nm $\leq \lambda \leq$ 500 nm), the transmittance increases smoothly and in the second region (200 nm $\leq \lambda \leq$ 240 nm) is the absorption region where the transmittance falls abruptly. Also, in the tertiary region (190 nm $\leq \lambda \leq$ 200 nm) the transmittances have an upsurge. The interference pattern in the transmittance manifests also absorbance the homogeneity of particle size. The results indicated Nanoparticle have high transmittance in the visible region. It is found out that average transmittance in the visible region is between 65%_ 80% at different concentrations, and maximum transmittance in this region is 88.5% for pure $SnO_2$ at 500 nm. Also, this result shows that the transmittance in the visible region decreases with increasing the Al percentage.

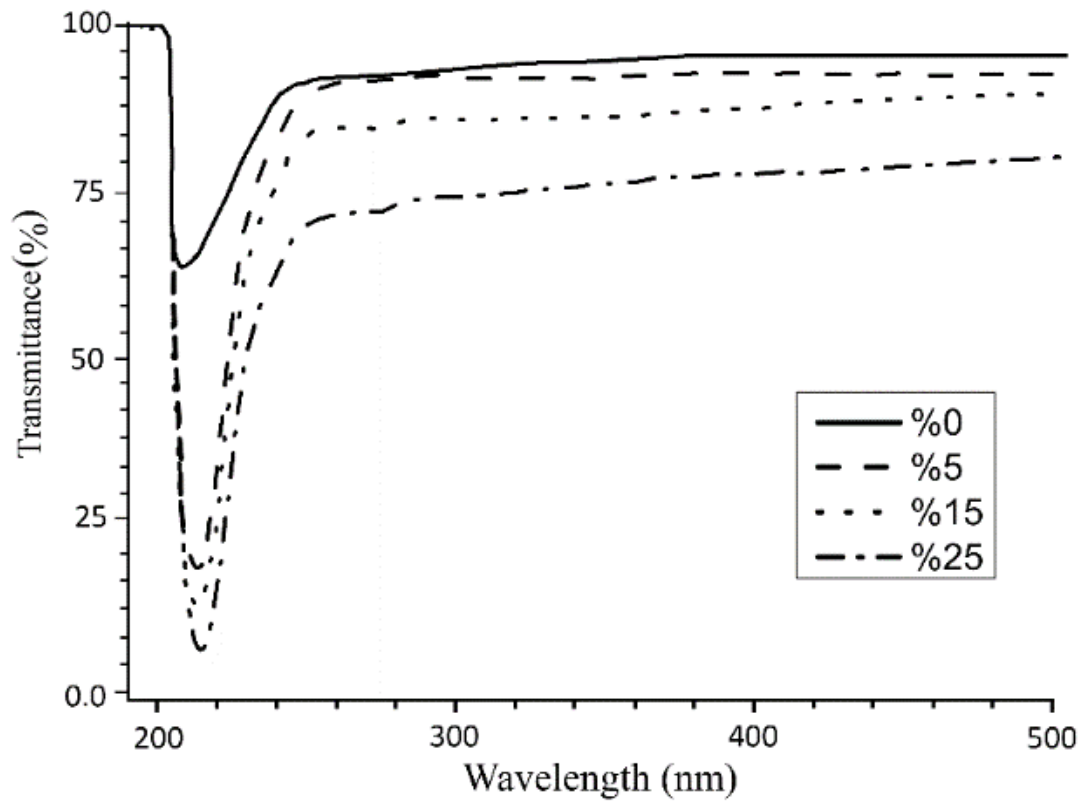

Fig. 4 UV-vis transmittance spectra of $SnO_2$ nanoparticles for Al-doped.

The absorption spectra showed in Fig. 5. As it can be seen, the nanoparticles have a low absorption in the first region, an upsurge in the second region, and in the tertiary region absorption fall abruptly. Transmittance spectrum is characterized by a sharp fall at wavelengths shorter than 250 nm, corresponding to the energy threshold for band edge absorption of $SnO_2$, thus transmittance fall and absorption abruptly in this region.

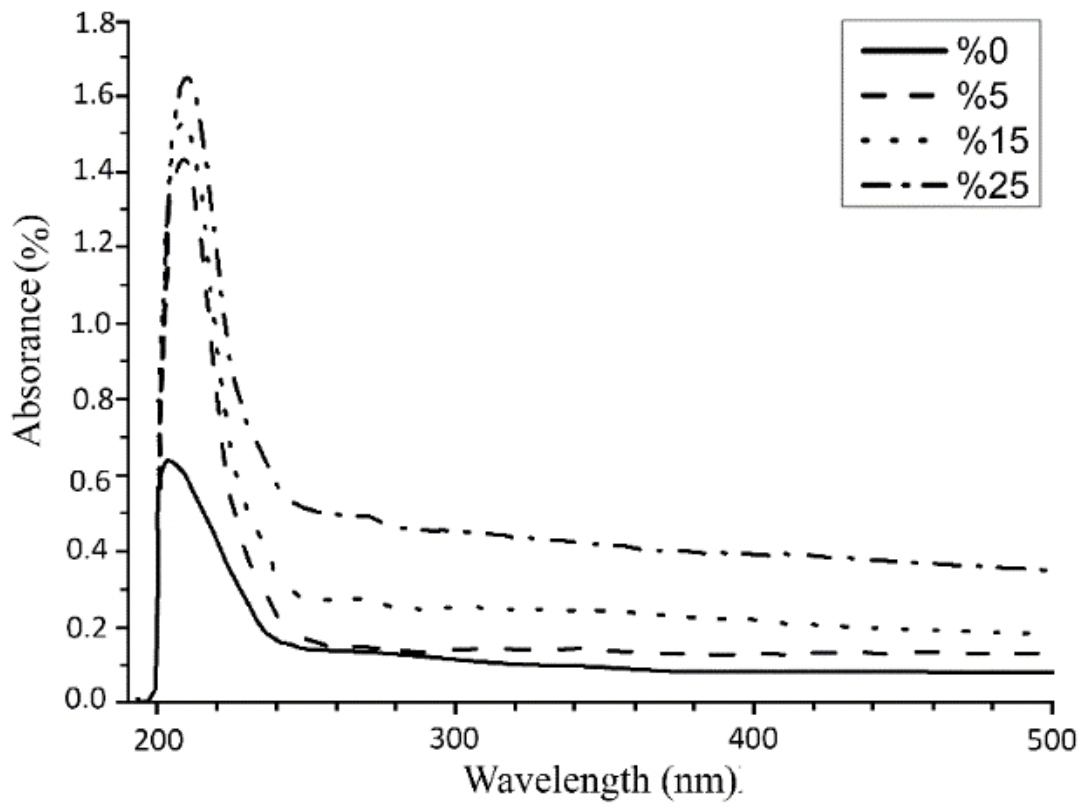

Fig. 5 UV-vis absorbance spectra of $SnO_2$ nanoparticles for Al-doped.

The edge of absorption got to shorter than wavelength when the Al-doped of the sol decreased, because with increasing the concentration of the sol the particle size become smaller, and smaller than particles will better absorbed the shorter wavelengths.

*The optical band gap*

The relation between the incident photon energy ($h\nu$) and the absorption coefficients ($\alpha$) is given by the following relation:

$$(\alpha h\nu)^{\frac{1}{n}} = A(h\nu - E_g) \qquad (3)$$

Here, A is a constant and $E_g$ the optical band gap of the material. The exponent (*n*) is dependent on the type of the optical transition. Note that quantity of (*n*) for direct allowed optical transition, indirect optical transition, and direct forbidden is equal to 1/2, 2, and 3/2, respectively [22, 23].

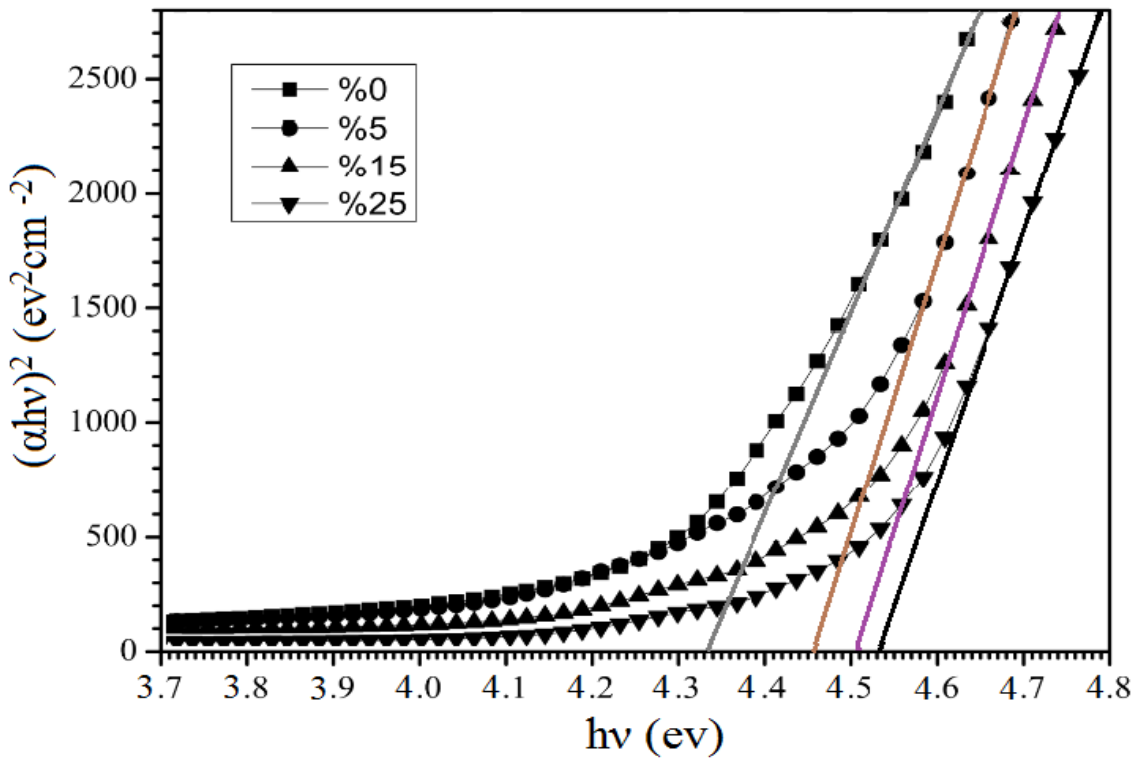

Fig. 6 Diagram of $SnO_2$ nanoparticles band gap Al-doped.

The Al-doped $SnO_2$ thin layers have a direct optical band. Therefore, the power (*n*) for them equals to 0.5 and it is possible to calculate the energy gap through the linear part of the curve at the zero absorption. We must plot a diagram $(\alpha h\nu)^2$ versus $h\nu$ at $\alpha = 0$. Fig. 6 illustrates the optical band gap for these nanoparticles.

The band gaps are equal to 4.33, 4.46, 4.51, and 4.53eV for pure, 5%, 15%, and 25% Al doped tin oxide, respectively (see Table 2). The direct band gap values of the nanoparticles increase with the increase of Al doping.

Table 2. Nanoparticles optical band gap with difference Al content

| Al –doped | Band gap energy (eV) |
|---|---|
| 0% | 4.33 |
| 5% | 4.46 |
| 15% | 4.51 |
| 25% | 4.53 |

The UV-visible absorbance shifts to the small wavelength when the doping Al decreases because with decreasing Al content, the particle size also reduces. Also, the small particles can better absorb at shorter wavelengths and with the decrease in size of the particles, the band gap increases. This is in full agreement with the analysis of XRD and SEM.

*FT-IR Analysis*

Figs. 7 and 8 show FT-IR spectra (FT-IR: Perkin Elmer BX (II)) of pure and Al-doped $SnO_2$ nanoparticles, respectively. Minimum transmittance on the wavelength $3400 - 3500\ cm^{-1}$ is related to the O-H bond stretching vibration of water molecules adsorbed [24].

Minimum transmittance in the range of $2300 - 2400\ cm^{-1}$ is caused by carbon dioxide. So by exposing the atmosphere and absorb carbon dioxide molecules are created [24].

The wavelength ranges between $1600 - 1700\ cm^{-1}$ is related to Flexural vibrations groups of O-H and in water molecules and Sn-OH bond [24].

Minimum transmittance in the range of $400 - 700\ cm^{-1}$ cause's vibration of Sn-O-Sn and Sn-O in $SnO_2$ molecule, and in wavelength $978 cm^{-1}$ related to Al-O bound [25-28].

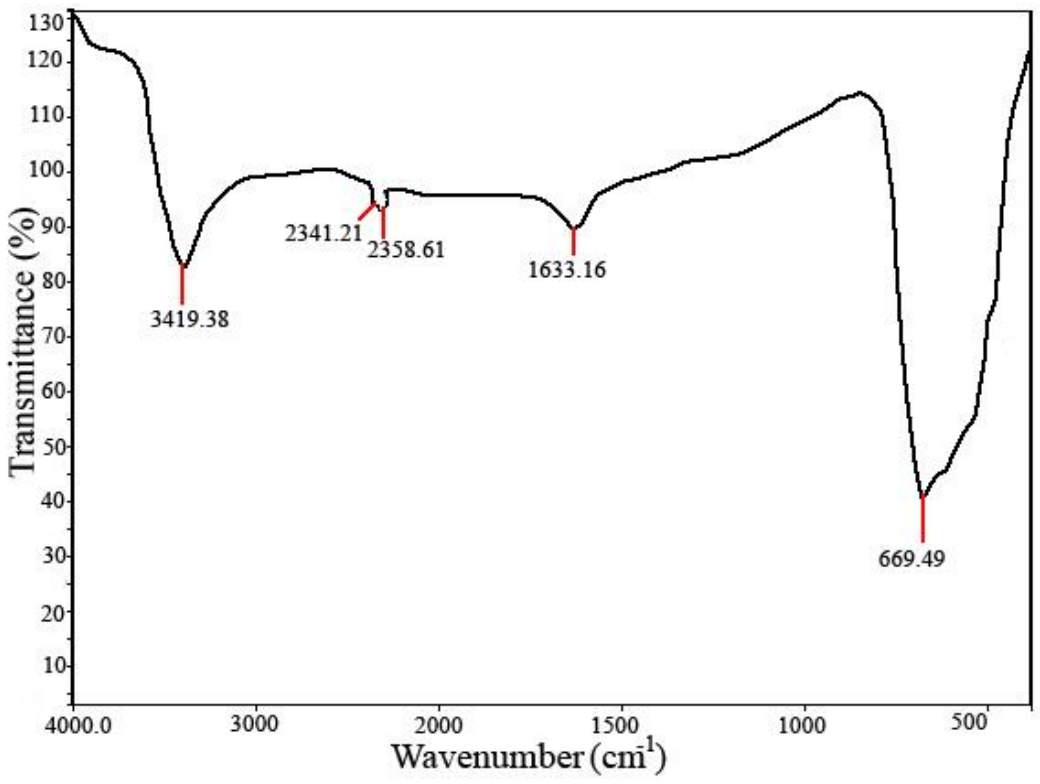

Fig. 7 FT-IR spectra of $SnO_2$ Nanoparticlespure

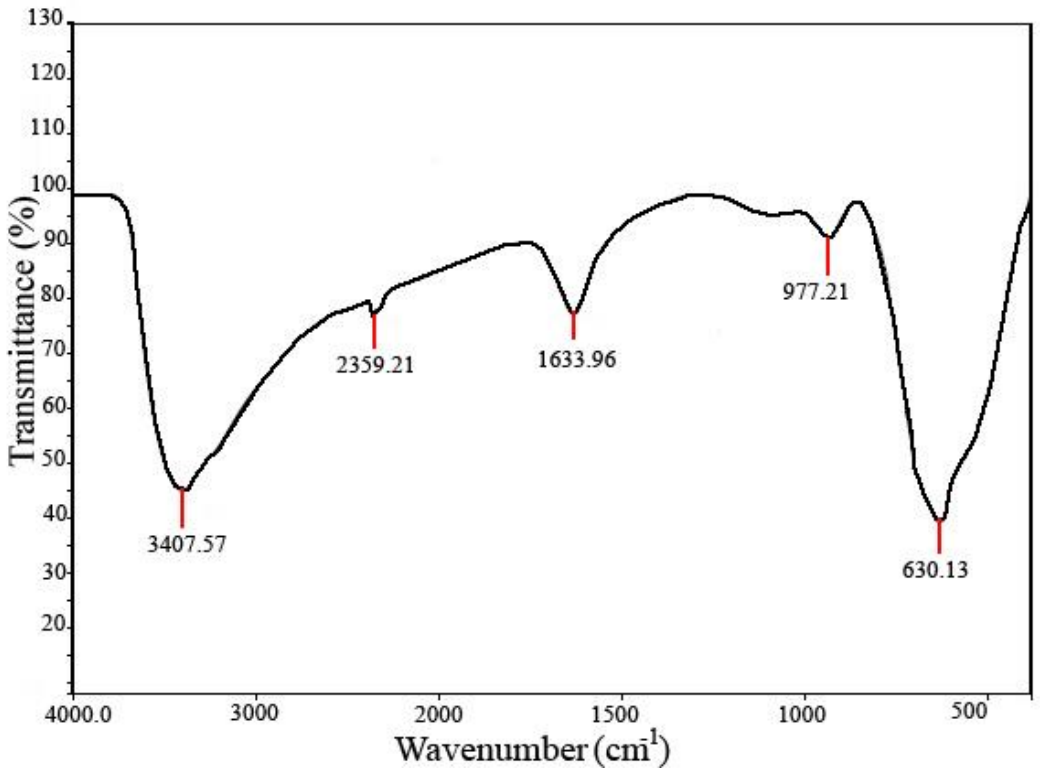

Fig. 8 FT-IR spectra of $SnO_2$Nanoparticle with Al-doped (25%)

*EDX Analysis*

Energy dispersive X-ray analysis (EDX) (Energy electron diffraction spectrometer (EDX)) is a technique for investigating the chemical composition of particles and also energy decomposition. Figs 9 and 10 show EDX analysis for pure and Al-doped $SnO_2$.

The EDX results in Fig. 9 show the presence of tin and Oxygen in the pure sample. The Al element is also observed in the doped sample via Fig. 10.

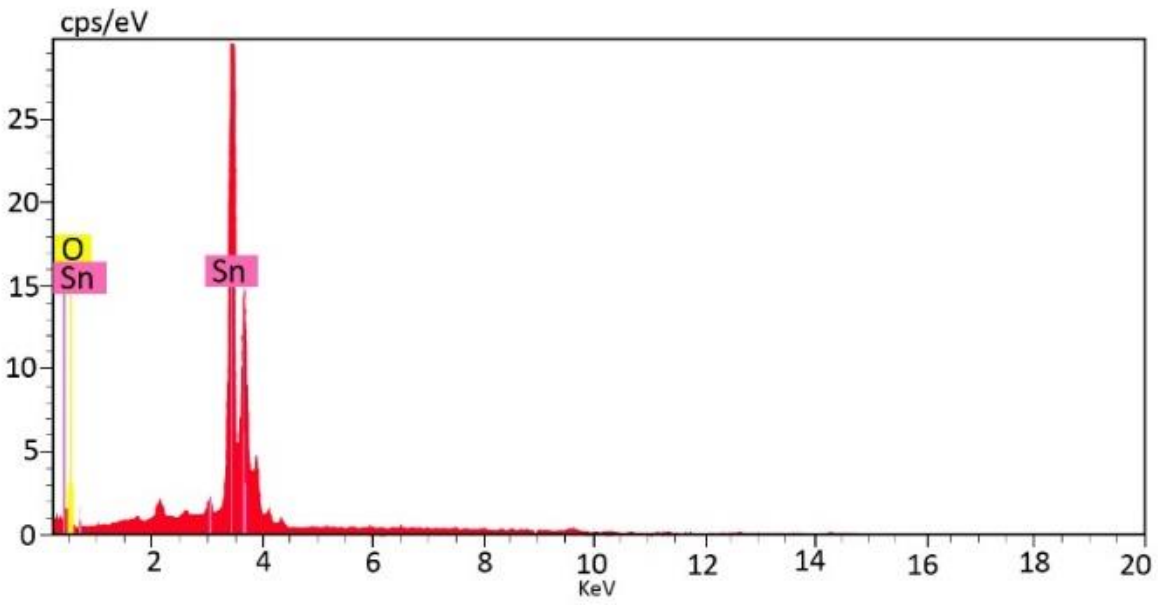

Fig. 9 EDX analysis of $SnO_2$ nanoparticle

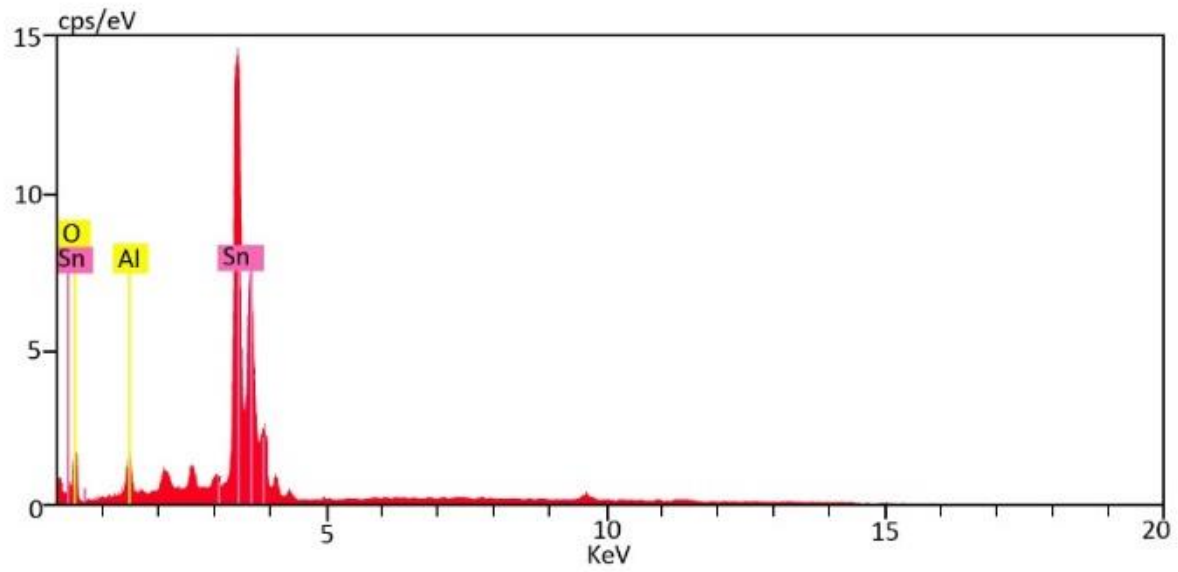

Fig. 10 EDX Analysis of $SnO_2$ Nanoparticle with Al-doped (25%)

CONCLUSION

Different percentages of Al-doped on $SnO_2$ nanoparticles were analyzed in this article. XRD diffraction showed that the particles are in tetragonal phase. Their intensity and crystallite size of the particles decrease by increase Al dopant.

SEM images show that average of particle size is about 18 nm, and the grain size decreased when the Al percent in the nanoparticles increased. The optical transmittance showed that nanoparticles have a high transmittance in the visible region. The energy band gap of the nanoparticles increases by increasing the Al percent value.

Optical analysis of UV-visible indicates that the energy gap increases by increasing the amount of Al doping. The band edge absorption of $SnO_2$ nanoparticles goes to the short wavelengths when the percent of the Al doped increased, because by increased the Al percent in the sol leads to smaller nanoparticles size, and smaller particles lead to better absorption at shorter wavelengths.

EDX analysis of the samples showed the existence of tin and Oxygen in the pure sample and also Al in the doped sample.

# REFRENCE

[1]. Vincent. C. A., The nature of semi conductivity in polycrystalline tin oxide, *Journal of The Electrochemical Society*, **119**, 515-518 (1972).

[2]. Chopra, K.L., Major, S. and Pandya, D.K., Transparent conductors—a status review. *Thin solid films*, **102**(1), 1-46 (1983).

[3]. Chen, Z., Lai, J.K.L., Shek, C.H. and Chen, H., Synthesis and structural characterization of rutile $SnO_2$ nanocrystals. *Journal of materials research*, **18**(06), 1289-1292 (2003).

[4]. Cirera, A., Vila, A., Dieguez, A., Cabot, A., Cornet, A. and Morante, J.R., Microwave processing for the low cost, mass production of undoped and in situ catalytic doped nanosized $SnO_2$ gas sensor powders. *Sensors and Actuators B: Chemical*, **64**(1), 65-69 (2000).

[5]. Liu, Y., Jiao, Y., Zhang, Z., Qu, F., Umar, A. and Wu, X., Hierarchical $SnO_2$ nanostructures made of intermingled ultrathin nanosheets for environmental remediation, smart gas sensor, and supercapacitor applications. *ACS applied materials & interfaces*, **6**(3), 2174-2184 (2014).

[6]. Suematsu, K., Shin, Y., Hua, Z., Yoshida, K., Yuasa, M., Kida, T. and Shimanoe, K., 2014. Nanoparticle cluster gas sensor: controlled clustering of $SnO_2$ nanoparticles for highly sensitive toluene detection. *ACS applied materials & interfaces*, **6**(7), 5319-5326 (2014).

[7]. Huu, N.K., Son, D.Y., Jang, I.H., Lee, C.R. and Park, N.G., Hierarchical $SnO_2$ nanoparticle-ZnO nanorod photoanode for improving transport and life time of photoinjected electrons in dye-sensitized solar cell. *ACS applied materials & interfaces*, **5**(3), 1038-1043 (2013).

[8]. Manjula, P., Boppella, R. and Manorama, S.V., A facile and green approach for the controlled synthesis of porous SnO2 nanospheres: application as an efficient photocatalyst and an excellent gas sensing material. *ACS applied materials & interfaces*, **4**(11), 6252-6260 (2012).

[9]. Pang, H., Yang, H., Guo, C.X. and Li, C.M., Functionalization of $SnO_2$ photoanode through Mg-doping and TiO2-coating to synergically boost dye-sensitized solar cell performance. *ACS applied materials & interfaces,* **4**(11), 6261-6265 (2012).

[10]. Wang, Y., Zhao, J.C., Zhang, S., Liu, Q.J. and Wu, X.H., Synthesis and optical properties of tin oxide nanocomposite with the ordered hexagonal mesostructure by mixed surfactant templating route. *Journal of non-crystalline solids*, **351**(16), 1477-1480 (2005).

[11]. Paraguay-Delgado, F., Antúnez-Flores, W., Miki-Yoshida, M., Aguilar-Elguezabal, A., Santiago, P., Diaz, R. and Ascencio, J.A., Structural analysis and growing mechanisms for long $SnO_2$ nanorods synthesized by spray pyrolysis. *Nanotechnology*, **16**(6), 688 (2005).

[12]. Geraldo, V., Scalvi, L.V.D.A., Morais, E.A.D., Santilli, C.V. and Pulcinelli, S.H., Sb doping effects and oxygen adsorption in SnO2 thin films deposited via sol-gel. *Materials Research*, **6**(4), 451-456 (2003).

[13]. Mishra, S., Ghanshyam, C., Ram, N., Singh, S., Bajpai, R.P. and Bedi, R.K., Alcohol sensing of tin oxide thin film prepared by sol-gel process. *Bulletin of Materials Science*, **25**(3), 231-234 (2002).

[14]. Kose, H., Aydin, A.O., Akbulut, H., The Effect of Temperature on Grain Size of SnO2 Nanoparticles Synthesized by Sol Gel Method, *ACTA PHYSICA POLONICA* ***A***, **125** (2) 345-347(2014).

[15]. Tripathy, S.K. and Hota, B.P., Influence of the Substrates Nature on Optical and Structural Characteristics of SnO2 Thin Film Prepared by Sol-Gel Technique. *Journal of Nano-and Electronic Physics*, **5**(3), 3012-1 (2013).

[16]. Gaber, A., Abdel-Rahim, M.A., Abdel-Latief, A.Y. and Abdel-Salam, M.N., Influence of calcination temperature on the structure and porosity of nanocrystalline $SnO_2$ synthesized by a conventional precipitation method. *Int J Electrochem Sci*, **9**(1), 81-95 (2014).

[17]. Goswami, Y.C., Kumar, V., Rajaram, P., Ganesan, V., Malik, M.A. and O'Brien, P., Synthesis of $SnO_2$ nanostructures by ultrasonic-assisted sol–gel method. *Journal of sol-gel science and technology*, **69**(3), 617-624 (2014).

[18]. Kim, M., Marom, N., Bobbitt, S. and Chelikowsky, J.R., The electronic and structural properties of SnO2 nanoparticles doped with antimony and fluorine. *In APS Meeting Abstracts***, 1** 44011 (2014).

[19]. Novinrooz, A., Sarabadani, P. and Garousi, J., Characterization of pure and antimony doped $SnO_2$ thin films prepared by the sol-gel technique. *Iranian Journal of Chemistry and Chemical Engineering (IJCCE)*, **25**(2), 31-38.(2006)

[20]. Gu, F., Wang, S.F., Lü, M.K., Zhou, G.J., Xu, D. and Yuan, D.R., Photoluminescence properties of $SnO_2$ nanoparticles synthesized by sol-gel method. *The Journal of Physical Chemistry* ***B***, **108**(24), 8119-8123 (2004).

[21]. Razeghizadeh, A., Elahi, E., Rafee, Investigation of UV-Vis Absorbance of $TiO_2$ Thin Films Sensitized with the Mulberry Pigment Cyanidin by Sol-Gel Method. *Nashrieh Shimi va Mohandesi Shimi Iran JA*, **35**(2), 1-8 (2016).

Pankove, J. I, *Optical Processes in Semiconductors* (Dover Publications Inc., New York), P. 1971, (1971)

[23]. Themlin, J.M., Sporken, R., Darville, J., Caudano, R., Gilles, J.M. and Johnson, R.L., Resonant-photoemission

study of SnO 2: cationic origin of the defect band-gap states. *Physical Review B*, **42**(18), 11914 (1990).

[24]. Brzyska, W., Spectral and thermal investigations of Y (III) and lanthanide (III) complexes with 3, 3-dimethylglutaric acid. *Polish Journal of Chemistry*, **75**(1), 43-47 (2001).

[25]. Nakamoto .K, *Infrared and Raman spectra of inorganic and coordination compounds*, 4th edn.( John Wiley & Sons, Inc) p 183, (1906)

[22]. hio, E, Busca, G, *Appl. Catal***B22**, 249(1999).

[26]. Gu Z, Liang P, Liu X, ZhangW, Le Y, *J Sol-Gel SciTechnol***10**, 159(2000).

[27]. Nakamoto K (ed), *Infrared spectra of inorganic and coordinated compounds*, (John Wiley & Sons, Inc), p 76, 06, (1963)

[28]. Gallardo-Amores, JM, Armaroli, T, Ramis, G, Finocc